\documentclass[%
 preprint,
 amsmath,amssymb,
 prl,
]{revtex4-2}
\usepackage{natbib}
\usepackage{graphicx}
\usepackage{dcolumn}
\usepackage{bm}

\begin{document}

\preprint{Preprint}

\title{Extreme low submergence in deep-canopy flows}

\author{Lo\"ic Chagot}
 \email{chagot@laplace.univ-tlse.fr}

 \affiliation{Institut de Mécanique des Fluides de Toulouse (IMFT), Université de Toulouse, Toulouse INP, UPS, CNRS, Toulouse, France}%
\affiliation{Laboratoire Plasma et Conversion d’énergie (LAPLACE), Université de Toulouse, Toulouse INP, UPS, CNRS, Toulouse, France}%

\author{Fr\'ed\'eric Y. Moulin}
 \email{frederic.moulin@imft.fr}
\affiliation{Institut de Mécanique des Fluides de Toulouse (IMFT), Université de Toulouse, Toulouse INP, UPS, CNRS, Toulouse, France 
}

\author{Olivier Eiff}
\email{olivier.eiff@kit.edu}
\affiliation{Institute for Water and Environment, Karlsruhe Institute
of Technology, Karlsruhe, Germany
}%

\date{\today}


\begin{abstract}
This letter investigates converged statistics in three-dimensional deep-canopy-dominated flows under two low relative submergence conditions: $h/k=1.5$ and $h/k=1.2$. Using a multi-plane telecentric PIV setup, time-averaged velocity fields were obtained across nine planes. For $h/k=1.5$, the flow structure exhibited a classical three-region behavior: a uniform emergent zone, a mixing zone and a logarithmic free-flow zone. In contrast, the extreme low submergence case ($h/k=1.2$) revealed significant flow modifications, including seiching—a periodic free-surface oscillation typically associated with emergent flows. Local double-averaged velocity profiles showed a unique undershoot near $z/k=0.8$, reflecting enhanced momentum transfer from cavity flows to alleys of the canopy. These findings highlight the transition mechanisms from submerged to emergent flow regimes, providing insights into canopy hydrodynamics and the influence of submergence on momentum transport.
\end{abstract}

\maketitle


\section{\label{sec:level1}Introduction:}

\indent

To describe river flows over rough beds, global parameters appearing in shallow water equations, bulk velocity and friction coefficient, are linked together using historical formulations (Chezy, Manning-Strickler, Darcy-Weisbach) \cite{wang2019friction,kadivar2021review,lee2024estimation}. A continuous evolution of the so-called friction laws from submerged to emergent flow conditions is assumed for flows over rough beds \cite{wang2019friction}.
Nowadays, the double-average methodology allows a better description and understanding of the vertical structure of flows over rough beds \cite{wilson1977,nikora2001,nikora2007,shounda2024rough}. Vertical profiles of double-averaged streamwise velocity and Reynolds shear stress yield, when integrated vertically, give access to the friction law for the considered flow regime \cite{li2020near}. Analysing vertical profiles of double-averaged quantities, \citet{nezu2008} showed that the flow for deep-canopy flows over aquatic submerged canopies separates into three regions : 1) a log-law region far above the canopy similar to the classic boundary layer over a rough bed, 2) a mixing region around the top of the canopy analogous to a pure plane mixing layer and 3) the lower canopy region, studied by \citep{nepf1999,nepf2000,tanino2008}, with the same structure as when the vegetation canopy is emergent (therefore also named emergent zone). 
Other authors simplified this three-region structure by splitting the flow into only two regions \citep{nepf2012,rubol2018}: above the canopy top, a logarithmic law is used to describe the velocity profile, whereas in the canopy below, a porous media model is used \citep{battiato2014}.

\indent
These simplified two-layer models assume the existence of a logarithmic law deep inside the roughness sublayer \citep[e.g.,][]{florens2013} and rely on continuity relationships at the top of the canopy to predict a velocity profile in accordance with the intermediate mixing layer discussed by \citet{nezu2008}. All these approaches are validated for strong levels of submergence, but are not verified for low submergence flow configurations.

\indent
Reliable velocity data within and near the canopy are scarce, in particular for low submergence flows. Studies \citep{manes2007,florens2013,rouzes2018} suggest that a logarithmic law behaviour in the flow just above the canopy exists and resists to the water level decrease for arrays of cubes ($h/k\geq 1.5$ where $h$ and $k$ are respectively the water level and the canopy height). However, \citet{rouzes2018} already observed that for $h/k=1.5$, the self-similarity of the turbulent boundary layer is lost, with an increase of the transverse turbulent fluctuations $\overline{v'^2}$ associated with a change in the shape of the turbulent eddies.

Based on robust telecentric-PIV measurements \citep{chagot2020}, this short communication  highlights how the decrease of the free surface towards the canopy top modifies strongly the structure of the flow for a very low submergence configuration just before transition to emergent flow configurations. This experimental observation challenges the idea of a continuous transition between submerged and emergent flow regimes.
To this end, a vegetation-like deep canopy of elongated cuboids in a square configuration is analysed. We show that the canopy flow structure changes abruptly before the transition between submergence and emergence, for a relative submergence $h/k=1.2$, with $h$ the water depth and $k$ the cuboid height. The double-averaged profiles are strongly modified, with the appearance of a non-monotonic velocity profile and seiching generation observed by \citep{dupuis2016} in emergent flow configurations. 

\section{Experimental set-up:}

\begin{figure}
\includegraphics[width=0.5\textwidth]{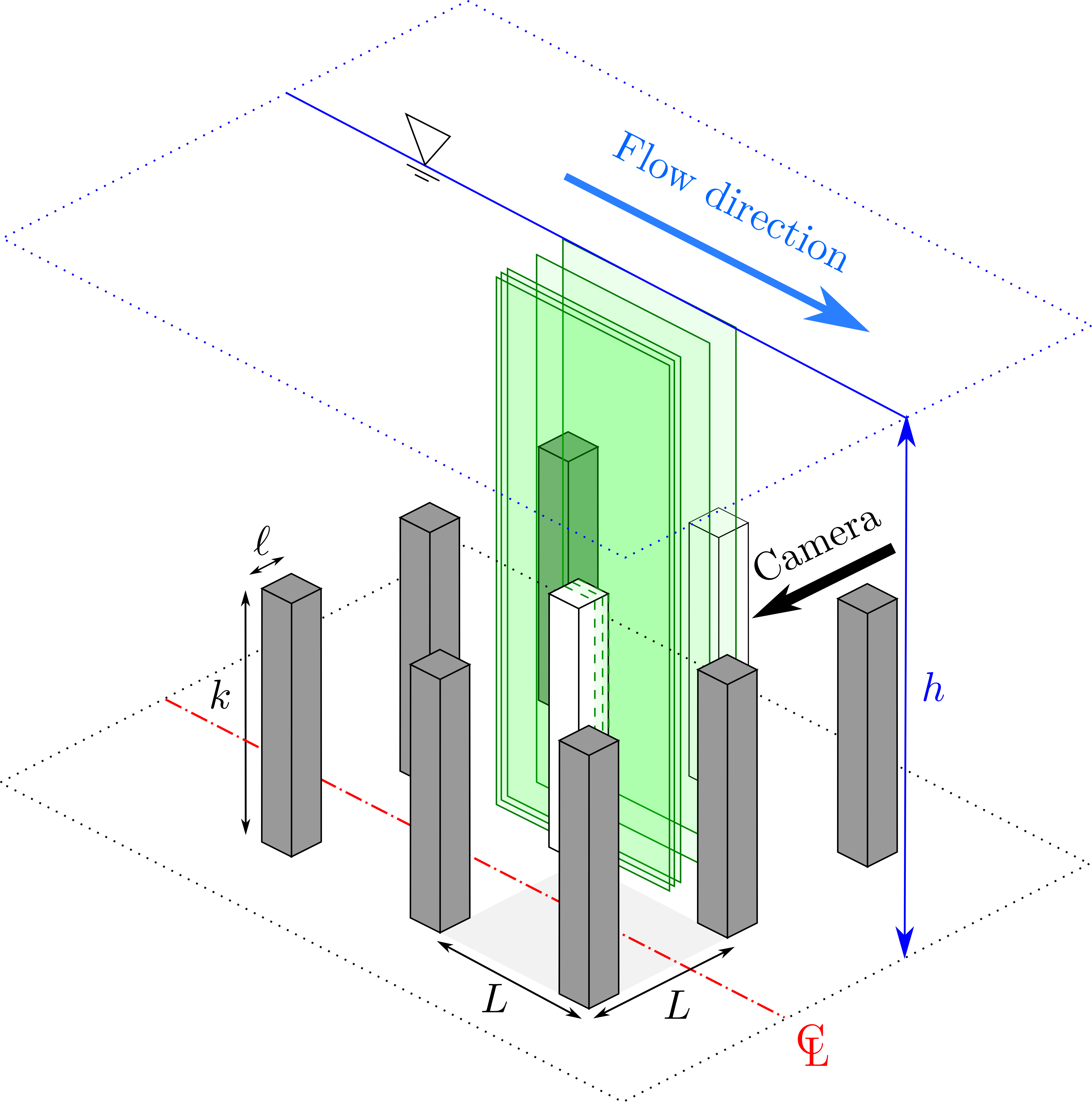}
\caption{\label{fig:exp_setup} Sketch of the telecentric PIV setup around the measurement area.}
\end{figure}

The experiments were performed at the Institut de M\'ecanique des Fluides de Toulouse (IMFT) in a 26 m long, 1.10 m wide ($W$) and 0.50 m deep open-channel flume made out of glass with a slope of 0.3 $\%$ \citep{rouzes2018}. To obtain uniform flow conditions with the fixed slope, the water level was adjusted with a downstream weir. The streamwise, transverse and wall-normal directions are defined as the $(x, y, z)$ coordinates. The origin of the coordinate system is taken at the end of the entry section,  at the bottom and centre of the flume.\\
\indent 
The deep canopy was built using cuboids with a square base of base length $\ell = 2$ cm  and height $k$ = 12 cm. A square configuration with a planar density $\lambda_p$ = 0.2 was chosen,  with $\lambda_p = \ell^2 /L^2$ and $L$ the distance between the cuboids (see Fig. \ref{fig:exp_setup}). Bed-porosity $\phi(z)$ with cuboids is independent of $z$, equal to $\phi=1-\lambda_p $ (here, $\phi$ = 0.8). To obtain fully developed flows, the cuboid bed extends along 5 m over $x=[16.7 - 21.7]$ m. The measurement section is located at $x_M=19.2 $ m.\\
\indent
In order to consider the effect of the very low relative submergence ($h/k$), two configurations were investigated : $h/k=1.5$ and 1.2. These parameters as well as other global flow parameters for the two beds studied are given in Table \ref{tab:table1}. It can be seen that both cases are sub-critical ($Fr < 1$).

\begin{table}
\caption{\label{tab:table1}Flow parameters for relative submergence ratio $h/k=$1.5 and 1.2. $h$: water depth; $\Phi=1-\lambda_p$: canopy porosity where $\lambda_p$ is the planar density; $U_b=Q/(h_e W)$: global bulk velocity where $W$ is the channel width $h_e=h-(1-\phi)k$ the effective water depth and $Q$ the water discharge; $u_k$: double-averaged streamwise velocity at $z=k$; $u_W$: wake velocity inferred from double-averaged streamwise velocity measurements in the emergent region; $Fr= U_b/\sqrt{g h_e}$: global Froude number.}
\begin{ruledtabular}
\begin{tabular}{lcr}
Parameter&$h/k=1.5$&$h/k=1.2$\\
\hline
$h$ (cm) & 18.2 & 14.0\\
$U_b$ (cm/s) & 18.8 & 9.6\\
$u_k$ (cm/s) & 20.4 & 17.9\\
$u_W$ (cm/s) & 7.9 & 8.5\\

$Fr$  & 0.15 & 0.09\\
\end{tabular}
\end{ruledtabular}
\end{table}

For data acquisition, the multi-plane telecentric PIV set-up built by \citet{chagot2020} was used here (see Fig. \ref{fig:exp_setup}). To get reliable uncorrelated statistics, 10000 image pairs were acquired at a frequency of 3 Hz (i.e., for 55.5 minutes) in each measurement plane. To achieve spatial convergence \citep{chagot2020}, the measurements were performed in five parallel vertical planes distributed along one central periodic pattern between the center of the cuboids and the center of the alley (see Fig. \ref{fig:exp_setup}). There are three planes located on the cuboids with a measurement step of $\Delta y=\ell/4$ and two planes in the alley with $\Delta y= (L-\ell)/4$. The transverse periodicity of the roughness pattern allows the introduction of four virtual planes by mirror symmetry. Double-averaged quantities presented here are the 9-planes double-averaged estimates as defined in \citet{chagot2020}. When the spatial averaging is performed only in one plane measurement, one-plane double-averaged quantities are obtained, following the notation of \citet{chagot2020}\\
\indent

For the extreme low submergence flow with $h/\ell=1.2$, a time-periodic deformation of the free surface was observed. Surprisingly, even if the Froude numbers are low for the two cases, respectively $Fr = 0.15$ and $0.09$ for $h/\ell=1.5$ and $h/\ell=1.2$ (see Table \ref{tab:table1}), a variation in time of the free surface ($\pm 0.5 cm$) was observed for $h/\ell=1.2$. This phenomenon, called a seiching, was observed by \citet{dupuis2016} for emerged cylinders in a square configuration. It is triggered by an interaction between the vortices shedded by the periodic pattern of cuboids on the one hand and the transverse gravity waves in the flume on the other hand. Such seiching was not observed for higher submergence flow conditions with $h/k \leq 1.5$. Its detection in our submerged experiment with $h/\ell=1.2$ reinforces the importance of this flow regime as an important step in the transition from submergence to emergence.  

\section{Results:}

\begin{figure*}
\includegraphics[width=0.8\textwidth]{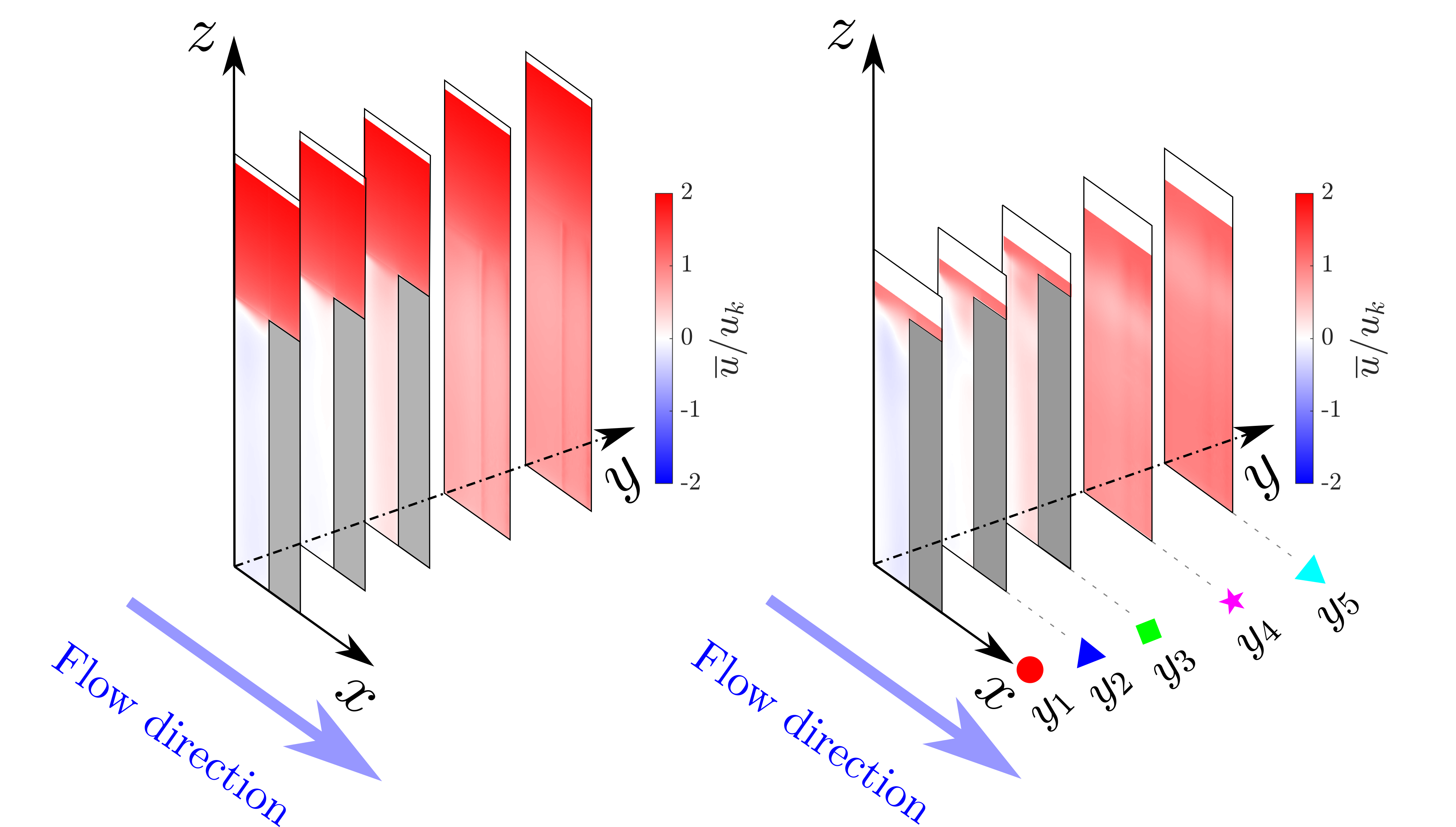}
\caption{\label{fig:3DFields} Streamwise mean velocity fields $\overline{u}$ in the different measurement planes for: $a)$ $h/k=1.5$ and $b)$ $h/k=1.2$.}
\end{figure*}
\indent
\subsection{Time-averaged streamwise velocity fields}
To analyse the flow structure modification, the mean streamwise velocity fields $\overline{u}$  are plotted in Fig. 2 in each PIV measurement plane, named $y_i$ with $i$ between $1$ and $5$.

\indent
 For $h/k=1.5$, a cavity flow is observed behind the cuboids (see planes $y_1$ and $y_2$) inside the canopy $z/k<1$, where the velocity $\overline{u}$ is negative. The cavity vortex vanishes at the cuboid edge (plane $y_3$) where $\overline{u}$ becomes positive. In the alley ($y_4$ and $y_5$), the velocity inside the canopy seems almost constant along the vertical.
Above the canopy ($z/k>1$), the free flow is very homogeneous in all horizontal directions and seems to be independent from the canopy flow below. Moreover, the transition layer between the two regions is very thin and located around the canopy top in all planes ($z/k=1$), even in the rear of the cuboids in planes $y_1$ and $y_2$. This flow structure for $h/k=1.5$ is in agreement with the two-region approach of \citet{rubol2018}.It also corresponds to a roughness sublayer extending not far away from the top of the canopy.

\indent
 For $h/k=1.2$, the flow behaviour is modified, especially near the top of cuboids. Behind the cuboids in planes $y_1$ and $y_2$, a significant penetration of the free flow in the canopy region is observed, with positive streamwise velocities (shown in red) plunging deep below the canopy top ($z/k<1$). More surprising is the apparition of a low velocity zone around $z/k \approx 0.8$ in the alley planes $y_4$ and $y_5$. Such a velocity deficit in the alley can only come from the low velocity region behind the cuboids. It indicates an increase of the momentum transversal exchange between the alley and the cavity flow in comparison with the previous reference case ($h/k=1.5$). 

\subsection{Double-averaged quantities}
\begin{figure*}
\includegraphics[width=1\textwidth]{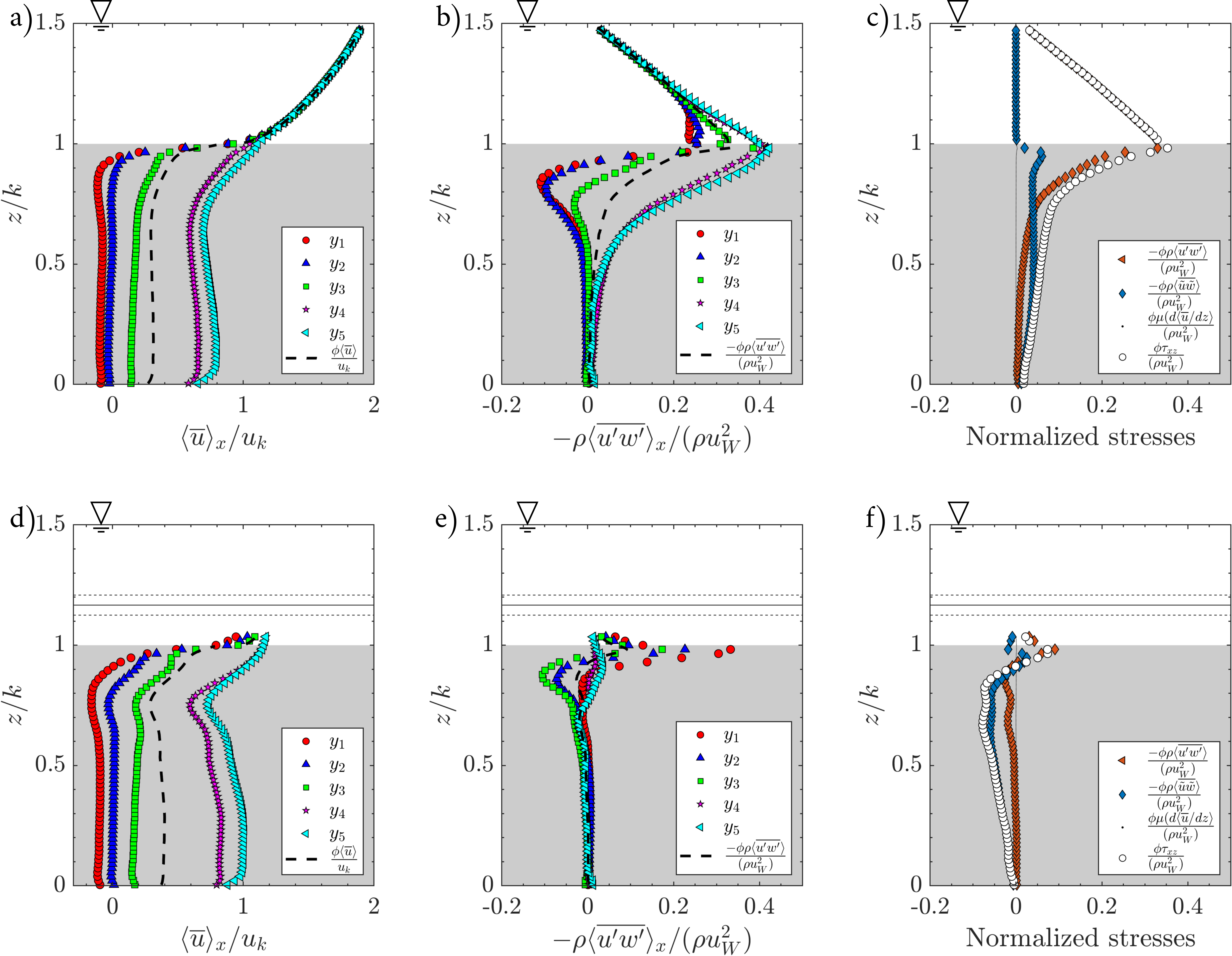}
\caption{\label{fig:DAx} Profiles of local double-averaged streamwise velocities ($\overline{u}/u_k$) for: $a)$ $h/k=1.5$ and $d)$ $h/k=1.2$. Profiles of local double-averaged Reynolds shear stresses ($-\phi\rho\overline{u'w'}/(\rho u_W^2)$) for: $b)$ $h/k=1.5$ and $e)$ $h/k=1.2$. Profiles of the normalized total shear stress $\phi\tau_{xz}/(\rho u_W^2)$ and the three contributing terms for: $c)$ $h/k=1.5$ and $f)$ $h/k=1.2$}
\end{figure*}

Fig. \ref{fig:DAx} shows the vertical profiles of the nine-plane double-averaged streamwise velocity ($\phi\langle\overline{u}\rangle$), of the Reynolds shear stress ($-\phi\rho\langle\overline{u'w'}\rangle$), and of the total shear stress ($-\phi\rho\langle\overline{u'w'}\rangle-\phi\rho\langle \tilde{u} \tilde{w} \rangle$). $u'$ and $w'$ are respectively the streamwise and wall-normal temporal fluctuations. $\tilde{u}$ and $\tilde{w}$ are respectively the streamwise and wall-normal dispersive terms. For the two configurations ($h/k=$ 1.5 and 1.2), Profiles of the one-plane double-averaged quantities calculated in each measurement plane ($\langle\overline{u}\rangle_x$, $-\rho\langle\overline{u'w'}\rangle_x$) and ($-\phi\rho\langle\overline{u'w'}\rangle_x-\phi\rho\langle \tilde{u} \tilde{w} \rangle_x$) are also plotted.

For $h/k=1.5$ (Fig. \ref{fig:DAx}-a,b), the classical structure of a uniform deep-canopy flow is obtained. As in \citet{nezu2008}, three regions are identified:
Above $z>1.1k$ the double-averaged streamwise velocity follows a logarithmic law. The double-averaged Reynolds shear stress $-\phi\rho\langle\overline{u'w'}\rangle$ is linear as expected for a 2D uniform flow without secondary currents, corresponding to a canonical rough-bed boundary layer. All one-plane double-averaged velocity profiles $\langle\overline{u}\rangle_x$ collapse on the double-averaged log-law profile for $z>1.1k$, indicating that this log-law zone is located just above the roughness sublayer. 
The mixing zone defined by \citet{nezu2008} is found here just below, in the region $0.65<z/k<1.1$. The upper limit $1.1$ corresponds to the top of the roughness sublayer. The lower limit $0.65$ corresponds to the transition to a constant value in the 9-planes double-averaged streamwise velocity profile.

The emergent zone is located below $z=0.65$ and corresponds to a constant value of the 9-planes double-averaged streamwise velocity, as expected from the equilibrium between the drag of cuboids and the flow forcing for a uniform flow. 

In this lower part of the flow (intensively studied by previous authors \citet{nepf1999,nepf2000,tanino2008}), the vertical transport of momentum should be completely negligible. It is the case here with a 9-planes double-averaged Reynolds tensor profile of Fig.\ref{fig:DAx}$-c$, where $-\phi\rho\langle\overline{u'w'}\rangle$ is lower than 10\% of its maximum found at the top of the cuboids $z/k=1$. 

In lower part of the emergent zone, $-\phi\rho\langle\overline{u'w'}\rangle_x$ reaches zero below $z/k<0.5$ in each measurement plane, as expected for a pure equilibrium between flow forcing and wake drag. It indicates that in the upper part of the emergent zone defined with the 9-planes double-averaged streamwise velocity profile, some vertical turbulent transport of momentum occurs in the region $0.5<z/k<0.65$.
Between $z/k=0.5$ and the top of the cuboids at $z/k=1$, a smooth evolution of the 9-planes double-averaged Reynolds stress profile $-\phi\rho\langle\overline{u'w'}\rangle$ is observed, as expected for the mixing region. However, individual behaviors of the 1-plane double-averaged Reynolds stress profiles are contrasted, especially when comparing planes behind the cuboids $y_1$, $y_2$ and $y_3$ and planes in the alley $y_4$ and $y_5$. In the alley, the 1-plane double-averaged Reynolds shear stress decreases linearly from a maximum at $z/k=1$ to $0$ around $z/k=0.5$. In the planes behind the cuboids, because of the cavity flow vortex, the 1-plane double-averaged Reynolds shear stress $-\rho\langle\overline{u'w'}\rangle_x$ becomes negative, with a minimum at the level of the vortex core, before returning to $0$ near $z/k=0.5$.

However, this flow structure is strongly modified for the very low relative submergence flow with $h/k=1.2$ (Fig. 4-b-d). The upper part of the free flow region (near $z=h$) was not accessible due to seiching. Yet, available data in the lower part shows the disappearance near the top of the cuboids of the collapse of all 1-plane double-averaged velocity profiles on a unique logarithmic velocity profile that was observed in the flow with $h/k=1.5$. The spatial dispersion, as defined by \citet{florens2013}, increases strongly and the top of the roughness sublayer is moved higher above the cuboids. Such a thickening of the roughness sublayer for low submergence was already observed by \citet{rouzes2018} for flow above cubes. Surprisingly, the 9-planes double-averaged velocity profile $\langle\overline{u}\rangle$ inside this thickening roughness sublayer still follows a logarithmic law. However, this log law behaviour loses all the similarity properties expected from the classical asymptotic overlapping theory between external and internal flow regions. Additionally, above the cuboids, the 9-planes double-averaged Reynolds shear stress profile is not linear anymore, in contrast with the flow $h/k=1.5$, and the determination of the shear velocity using the slope of $-\rho\langle\overline{u'w'}\rangle$ is no longer possible.

In the emergent zone, below $z/k<0.65$, the 1-plane double-averaged velocity profiles follow the same behaviour as for $h/k=1.5$. Yet, in the alley for planes $y_4$ and $y_5$, the lowering of the free surface enhances the convex shape of the velocity profiles. This convex shape in the alley reminds the one of velocity profiles in narrow channels (see \citet{Oukacine2022}), suggesting that for very low submergence levels, alleys act as independent parallel channels in the emergent region.

The most striking impact of the lowering free surface on the flow structure is found in the exchange region.  A strong dip of velocity is observed around $z/k = 0.8$ in all 1-plane double-averaged velocity profiles, with values lower than the constant one found in the emergent region below. Time-averaged velocity fields in Fig. \ref{fig:3DFields}-b show that this drop of the 1-plane double-averaged streamwise velocity in each measurement plane is a velocity decrease that occupies the whole periodic pattern in the main flow direction. The position of the drop of velocity corresponds also to a well identified secondary current pattern located in the mixing region, which transports a deficit of velocity from the the cavity vortex flow behind the cuboids into the alley. The significant penetration of the free flow by the canopy top in planes $y_1$ and $y_2$ for this flow is compensated by  a transverse flow from the cuboids to the alleys around $z/k=0.8$
In the mixing region for flow $h/k=1.2$, 1-plane double-averaged Reynolds shear stress profiles in Fig. 4d all drop very quickly to zero in all measurement planes. This faster drop than for $h/k=1.5$ indicates that for the extreme low submergence configuration with $h/k=1.2$, all the turbulent vertical transport of momentum is killed. 

\section{Discussion}

Using accurate telecentric 2D PIV measurements \citep{chagot2020} on a periodic elongated canopy, an extreme low submergence configuration ($h/k=1.2$) was investigated and compared with a more regular flow structure found for a low submergence configuration ($h/k=1.5$) and well discussed in previous studies \citep{florens2013,rouzes2018,nezu2008}. 

The most obvious phenomena that distinguises the extreme low submergence configuration ($h/k=1.2$) from a regular low submergence flow configuration ($h/k \le 1.5$) is the occurence of a periodic free-surface fluctuation know as seiching, that was observed only in emergent configurations until now. Indeed, in regular low submergence flow configurations, a turbulent free flow layer exists above the roughness sublayer and transports turbulent fluctuations both horizontally and vertically through the canopy top, preventing any locking mechanism between organised cavity flows behind the cuboids and a large scale free surface oscillation. In contrast,   seiching  occurs  in emergent configurations because  cavity flows and vortex shedding behind the cuboids sychronize with  available frequencies of transverse free surface waves. Here, the occurence of a seiching for the extreme low submergence flow ($h/k=1.2$) indicates that its properties are closer to those of emergent than submerged flow configurations.    

The analysis of statistics for the streamwise velocity and the Reynolds stress confirms the disappearance of the horizontally homogeneous turbulent free flow layer. This disappearance is associated with the thickening of the roughness sublayer and the massive decrease of the Reynolds stress in the mixing zone around the top of the cuboids. Indeed, all turbulent fluxes of momentum between the different part of the flow (cavity, alleys and free flow above the top of the cubes) are killed. Secondary currents in cross-sectional plane $(y,z)$ perpendicular to the main flow direction are then the only remaining mechanism for momentum transport. These secondary currents were already identified and discussed by \citet{rouzes2018} in low submergence flows above cubes, but with experimental data available only above the top of the cubes. In the present investigation, we observe that these secondary currents are aligned along the edges of the cuboids and coincide with the mixing region. For extreme low submergence flow conditions ($h/k=1.2$), just before the transition to emergent flows, they are still present and completely dominate all momentum fluxes in the mixing region since turbulent fluxes of momentum are all collapsing. The most striking consequence is an increased transport of momentum around $z/k=0.8$ from the cavity flow behind the cuboids towards the alleys, leading to the strong drop of velocity observed around $z/k=0.8$ in the alley. When horizontally averaged in the framework of the double-averaging methodology, it generates a very exotic profile for the double-averaged streamwise velocity : the monotonic decrease of the velocity from its log-law behaviour in the free flow zone to its constant value in the emergent zone is replaced by an undershoot towards a minimum just below the top of the rough elements (here $z/k=0.8$) before its return to an higher constant value in the emergent zone. In the emergent zone, the 1-plane double-averaged velocity profiles inside the alley become similar to the velocity profiles observed by \citet{Oukacine2022} in emergent flows with cubes, exhibiting a flow structure very close to the one in narrow channels.

In conclusion, the transition from  submerged to emergent flows is far from smooth : just before reaching emergent flow conditions, exotic vertical velocity profiles with undershooting below the top of the roughness elements are generated, forced by persistant secondary currents located near the top of the canopy that finally control momentum exchanges there just before the transition. In the meantime, these extreme low submergence flows already share common properties with emergent flows : they are prone to the seiching phenonemon and exhibit the flow channelisation observed and discussed by \citet{Oukacine2022}.  

\begin{acknowledgments}
This work was funded by the French National
Research Agency (Flowres project, Grant no. ANR-14-CE03-0010) and
was performed using HPC resources from CALMIP (project P16012).
\end{acknowledgments}

\bibliography{apssamp.bib}

\providecommand{\noopsort}[1]{}\providecommand{\singleletter}[1]{#1}%
\begin{thebibliography}{21}%
\makeatletter
\providecommand \@ifxundefined [1]{%
 \@ifx{#1\undefined}
}%
\providecommand \@ifnum [1]{%
 \ifnum #1\expandafter \@firstoftwo
 \else \expandafter \@secondoftwo
 \fi
}%
\providecommand \@ifx [1]{%
 \ifx #1\expandafter \@firstoftwo
 \else \expandafter \@secondoftwo
 \fi
}%
\providecommand \natexlab [1]{#1}%
\providecommand \enquote  [1]{``#1''}%
\providecommand \bibnamefont  [1]{#1}%
\providecommand \bibfnamefont [1]{#1}%
\providecommand \citenamefont [1]{#1}%
\providecommand \href@noop [0]{\@secondoftwo}%
\providecommand \href [0]{\begingroup \@sanitize@url \@href}%
\providecommand \@href[1]{\@@startlink{#1}\@@href}%
\providecommand \@@href[1]{\endgroup#1\@@endlink}%
\providecommand \@sanitize@url [0]{\catcode `\\12\catcode `\$12\catcode
  `\&12\catcode `\#12\catcode `\^12\catcode `\_12\catcode `\%12\relax}%
\providecommand \@@startlink[1]{}%
\providecommand \@@endlink[0]{}%
\providecommand \url  [0]{\begingroup\@sanitize@url \@url }%
\providecommand \@url [1]{\endgroup\@href {#1}{\urlprefix }}%
\providecommand \urlprefix  [0]{URL }%
\providecommand \Eprint [0]{\href }%
\providecommand \doibase [0]{https://doi.org/}%
\providecommand \selectlanguage [0]{\@gobble}%
\providecommand \bibinfo  [0]{\@secondoftwo}%
\providecommand \bibfield  [0]{\@secondoftwo}%
\providecommand \translation [1]{[#1]}%
\providecommand \BibitemOpen [0]{}%
\providecommand \bibitemStop [0]{}%
\providecommand \bibitemNoStop [0]{.\EOS\space}%
\providecommand \EOS [0]{\spacefactor3000\relax}%
\providecommand \BibitemShut  [1]{\csname bibitem#1\endcsname}%
\let\auto@bib@innerbib\@empty
\bibitem [{\citenamefont {Wang}\ \emph {et~al.}(2019)\citenamefont {Wang},
  \citenamefont {Peng}, \citenamefont {Huai}, \citenamefont {Katul},
  \citenamefont {Liu}, \citenamefont {Qu},\ and\ \citenamefont
  {Dong}}]{wang2019friction}%
  \BibitemOpen
  \bibfield  {author} {\bibinfo {author} {\bibfnamefont {W.-J.}\ \bibnamefont
  {Wang}}, \bibinfo {author} {\bibfnamefont {W.-Q.}\ \bibnamefont {Peng}},
  \bibinfo {author} {\bibfnamefont {W.-X.}\ \bibnamefont {Huai}}, \bibinfo
  {author} {\bibfnamefont {G.~G.}\ \bibnamefont {Katul}}, \bibinfo {author}
  {\bibfnamefont {X.-B.}\ \bibnamefont {Liu}}, \bibinfo {author} {\bibfnamefont
  {X.-D.}\ \bibnamefont {Qu}},\ and\ \bibinfo {author} {\bibfnamefont
  {F.}~\bibnamefont {Dong}},\ }\bibfield  {title} {\bibinfo {title} {Friction
  factor for turbulent open channel flow covered by vegetation},\ }\href@noop
  {} {\bibfield  {journal} {\bibinfo  {journal} {Scientific reports}\ }\textbf
  {\bibinfo {volume} {9}},\ \bibinfo {pages} {5178} (\bibinfo {year}
  {2019})}\BibitemShut {NoStop}%
\bibitem [{\citenamefont {Kadivar}\ \emph {et~al.}(2021)\citenamefont
  {Kadivar}, \citenamefont {Tormey},\ and\ \citenamefont
  {McGranaghan}}]{kadivar2021review}%
  \BibitemOpen
  \bibfield  {author} {\bibinfo {author} {\bibfnamefont {M.}~\bibnamefont
  {Kadivar}}, \bibinfo {author} {\bibfnamefont {D.}~\bibnamefont {Tormey}},\
  and\ \bibinfo {author} {\bibfnamefont {G.}~\bibnamefont {McGranaghan}},\
  }\bibfield  {title} {\bibinfo {title} {A review on turbulent flow over rough
  surfaces: Fundamentals and theories},\ }\href@noop {} {\bibfield  {journal}
  {\bibinfo  {journal} {International Journal of Thermofluids}\ }\textbf
  {\bibinfo {volume} {10}},\ \bibinfo {pages} {100077} (\bibinfo {year}
  {2021})}\BibitemShut {NoStop}%
\bibitem [{\citenamefont {Lee}\ \emph {et~al.}(2024)\citenamefont {Lee},
  \citenamefont {Park}, \citenamefont {Lee}, \citenamefont {Song},\ and\
  \citenamefont {Park}}]{lee2024estimation}%
  \BibitemOpen
  \bibfield  {author} {\bibinfo {author} {\bibfnamefont {M.}~\bibnamefont
  {Lee}}, \bibinfo {author} {\bibfnamefont {Y.~S.}\ \bibnamefont {Park}},
  \bibinfo {author} {\bibfnamefont {M.}~\bibnamefont {Lee}}, \bibinfo {author}
  {\bibfnamefont {Y.-S.}\ \bibnamefont {Song}},\ and\ \bibinfo {author}
  {\bibfnamefont {C.}~\bibnamefont {Park}},\ }\bibfield  {title} {\bibinfo
  {title} {Estimation of friction factor and bed shear stress considering
  bedform effect in rivers},\ }\href@noop {} {\bibfield  {journal} {\bibinfo
  {journal} {Earth Surface Processes and Landforms}\ } (\bibinfo {year}
  {2024})}\BibitemShut {NoStop}%
\bibitem [{\citenamefont {Wilson}\ and\ \citenamefont
  {Shaw}(1977)}]{wilson1977}%
  \BibitemOpen
  \bibfield  {author} {\bibinfo {author} {\bibfnamefont {N.}~\bibnamefont
  {Wilson}}\ and\ \bibinfo {author} {\bibfnamefont {R.}~\bibnamefont {Shaw}},\
  }\bibfield  {title} {\bibinfo {title} {Higher-order closure model for canopy
  flow},\ }\href@noop {} {\bibfield  {journal} {\bibinfo  {journal} {J. of
  Applied Meteorology}\ }\textbf {\bibinfo {volume} {16}},\ \bibinfo {pages}
  {1197} (\bibinfo {year} {1977})}\BibitemShut {NoStop}%
\bibitem [{\citenamefont {Nikora}\ \emph {et~al.}(2001)\citenamefont {Nikora},
  \citenamefont {Goring}, \citenamefont {McEwan},\ and\ \citenamefont
  {Griffiths}}]{nikora2001}%
  \BibitemOpen
  \bibfield  {author} {\bibinfo {author} {\bibfnamefont {V.}~\bibnamefont
  {Nikora}}, \bibinfo {author} {\bibfnamefont {D.}~\bibnamefont {Goring}},
  \bibinfo {author} {\bibfnamefont {I.}~\bibnamefont {McEwan}},\ and\ \bibinfo
  {author} {\bibfnamefont {G.}~\bibnamefont {Griffiths}},\ }\bibfield  {title}
  {\bibinfo {title} {Spatially averaged open-channel flow over rough bed},\
  }\href@noop {} {\bibfield  {journal} {\bibinfo  {journal} {J. Hydraul. Eng}\
  }\textbf {\bibinfo {volume} {127}},\ \bibinfo {pages} {123} (\bibinfo {year}
  {2001})}\BibitemShut {NoStop}%
\bibitem [{\citenamefont {Nikora}\ \emph {et~al.}(2007)\citenamefont {Nikora},
  \citenamefont {McEwan}, \citenamefont {McLean}, \citenamefont {Coleman},
  \citenamefont {Pokrajac},\ and\ \citenamefont {Walters}}]{nikora2007}%
  \BibitemOpen
  \bibfield  {author} {\bibinfo {author} {\bibfnamefont {V.}~\bibnamefont
  {Nikora}}, \bibinfo {author} {\bibfnamefont {I.}~\bibnamefont {McEwan}},
  \bibinfo {author} {\bibfnamefont {S.}~\bibnamefont {McLean}}, \bibinfo
  {author} {\bibfnamefont {S.}~\bibnamefont {Coleman}}, \bibinfo {author}
  {\bibfnamefont {D.}~\bibnamefont {Pokrajac}},\ and\ \bibinfo {author}
  {\bibfnamefont {R.}~\bibnamefont {Walters}},\ }\bibfield  {title} {\bibinfo
  {title} {Double-averaging concept for rough-bed open-channel and overland
  flows: Theoretical background},\ }\href@noop {} {\bibfield  {journal}
  {\bibinfo  {journal} {J. Hydraul. Eng.}\ }\textbf {\bibinfo {volume} {133}},\
  \bibinfo {pages} {873} (\bibinfo {year} {2007})}\BibitemShut {NoStop}%
\bibitem [{\citenamefont {Shounda}\ \emph {et~al.}(2024)\citenamefont
  {Shounda}, \citenamefont {Barman},\ and\ \citenamefont
  {Debnath}}]{shounda2024rough}%
  \BibitemOpen
  \bibfield  {author} {\bibinfo {author} {\bibfnamefont {J.}~\bibnamefont
  {Shounda}}, \bibinfo {author} {\bibfnamefont {K.}~\bibnamefont {Barman}},\
  and\ \bibinfo {author} {\bibfnamefont {K.}~\bibnamefont {Debnath}},\
  }\bibfield  {title} {\bibinfo {title} {Rough bed hydrodynamics under
  wave-current interactional flow field: double averaging approach},\
  }\href@noop {} {\bibfield  {journal} {\bibinfo  {journal} {Journal of
  Turbulence}\ }\textbf {\bibinfo {volume} {25}},\ \bibinfo {pages} {247}
  (\bibinfo {year} {2024})}\BibitemShut {NoStop}%
\bibitem [{\citenamefont {Li}\ and\ \citenamefont {Li}(2020)}]{li2020near}%
  \BibitemOpen
  \bibfield  {author} {\bibinfo {author} {\bibfnamefont {J.}~\bibnamefont
  {Li}}\ and\ \bibinfo {author} {\bibfnamefont {S.~S.}\ \bibnamefont {Li}},\
  }\bibfield  {title} {\bibinfo {title} {Near-bed velocity and shear stress of
  open-channel flow over surface roughness},\ }\href@noop {} {\bibfield
  {journal} {\bibinfo  {journal} {Environmental Fluid Mechanics}\ }\textbf
  {\bibinfo {volume} {20}},\ \bibinfo {pages} {293} (\bibinfo {year}
  {2020})}\BibitemShut {NoStop}%
\bibitem [{\citenamefont {Nezu}\ and\ \citenamefont {Sanjou}(2008)}]{nezu2008}%
  \BibitemOpen
  \bibfield  {author} {\bibinfo {author} {\bibfnamefont {I.}~\bibnamefont
  {Nezu}}\ and\ \bibinfo {author} {\bibfnamefont {M.}~\bibnamefont {Sanjou}},\
  }\bibfield  {title} {\bibinfo {title} {Turburence structure and coherent
  motion in vegetated canopy open-channel flows},\ }\href@noop {} {\bibfield
  {journal} {\bibinfo  {journal} {J Hydro-Environ Res}\ }\textbf {\bibinfo
  {volume} {2}},\ \bibinfo {pages} {62} (\bibinfo {year} {2008})}\BibitemShut
  {NoStop}%
\bibitem [{\citenamefont {Nepf}(1999)}]{nepf1999}%
  \BibitemOpen
  \bibfield  {author} {\bibinfo {author} {\bibfnamefont {H.~M.}\ \bibnamefont
  {Nepf}},\ }\bibfield  {title} {\bibinfo {title} {Drag, turbulence, and
  diffusion in flow through emergent vegetation},\ }\href@noop {} {\bibfield
  {journal} {\bibinfo  {journal} {Water resources research}\ }\textbf {\bibinfo
  {volume} {35}},\ \bibinfo {pages} {479} (\bibinfo {year} {1999})}\BibitemShut
  {NoStop}%
\bibitem [{\citenamefont {Nepf}\ and\ \citenamefont {Vivoni}(2000)}]{nepf2000}%
  \BibitemOpen
  \bibfield  {author} {\bibinfo {author} {\bibfnamefont {H.~M.}\ \bibnamefont
  {Nepf}}\ and\ \bibinfo {author} {\bibfnamefont {E.}~\bibnamefont {Vivoni}},\
  }\bibfield  {title} {\bibinfo {title} {Flow structure in depth-limited,
  vegetated flow},\ }\href@noop {} {\bibfield  {journal} {\bibinfo  {journal}
  {Journal of Geophysical Research: Oceans}\ }\textbf {\bibinfo {volume}
  {105}},\ \bibinfo {pages} {28547} (\bibinfo {year} {2000})}\BibitemShut
  {NoStop}%
\bibitem [{\citenamefont {Tanino}\ and\ \citenamefont
  {Nepf}(2008)}]{tanino2008}%
  \BibitemOpen
  \bibfield  {author} {\bibinfo {author} {\bibfnamefont {Y.}~\bibnamefont
  {Tanino}}\ and\ \bibinfo {author} {\bibfnamefont {H.~M.}\ \bibnamefont
  {Nepf}},\ }\bibfield  {title} {\bibinfo {title} {Laboratory investigation of
  mean drag in a random array of rigid, emergent cylinders},\ }\href@noop {}
  {\bibfield  {journal} {\bibinfo  {journal} {Journal of Hydraulic
  Engineering}\ }\textbf {\bibinfo {volume} {134}},\ \bibinfo {pages} {34}
  (\bibinfo {year} {2008})}\BibitemShut {NoStop}%
\bibitem [{\citenamefont {Nepf}(2012)}]{nepf2012}%
  \BibitemOpen
  \bibfield  {author} {\bibinfo {author} {\bibfnamefont {H.~M.}\ \bibnamefont
  {Nepf}},\ }\bibfield  {title} {\bibinfo {title} {Flow and transport in
  regions with aquatic vegetation},\ }\href@noop {} {\bibfield  {journal}
  {\bibinfo  {journal} {Annual review of fluid mechanics}\ }\textbf {\bibinfo
  {volume} {44}},\ \bibinfo {pages} {123} (\bibinfo {year} {2012})}\BibitemShut
  {NoStop}%
\bibitem [{\citenamefont {Rubol}\ \emph {et~al.}(2018)\citenamefont {Rubol},
  \citenamefont {Ling},\ and\ \citenamefont {Battiato}}]{rubol2018}%
  \BibitemOpen
  \bibfield  {author} {\bibinfo {author} {\bibfnamefont {S.}~\bibnamefont
  {Rubol}}, \bibinfo {author} {\bibfnamefont {B.}~\bibnamefont {Ling}},\ and\
  \bibinfo {author} {\bibfnamefont {I.}~\bibnamefont {Battiato}},\ }\bibfield
  {title} {\bibinfo {title} {Universal scaling-law for flow resistance over
  canopies with complex morphology},\ }\href@noop {} {\bibfield  {journal}
  {\bibinfo  {journal} {Scientific reports}\ }\textbf {\bibinfo {volume} {8}},\
  \bibinfo {pages} {1} (\bibinfo {year} {2018})}\BibitemShut {NoStop}%
\bibitem [{\citenamefont {Battiato}\ and\ \citenamefont
  {Rubol}(2014)}]{battiato2014}%
  \BibitemOpen
  \bibfield  {author} {\bibinfo {author} {\bibfnamefont {I.}~\bibnamefont
  {Battiato}}\ and\ \bibinfo {author} {\bibfnamefont {S.}~\bibnamefont
  {Rubol}},\ }\bibfield  {title} {\bibinfo {title} {Single-parameter model of
  vegetated aquatic flows},\ }\href@noop {} {\bibfield  {journal} {\bibinfo
  {journal} {Water Resour. Res.}\ }\textbf {\bibinfo {volume} {50}},\ \bibinfo
  {pages} {6358} (\bibinfo {year} {2014})}\BibitemShut {NoStop}%
\bibitem [{\citenamefont {Florens}\ \emph {et~al.}(2013)\citenamefont
  {Florens}, \citenamefont {Eiff},\ and\ \citenamefont {Moulin}}]{florens2013}%
  \BibitemOpen
  \bibfield  {author} {\bibinfo {author} {\bibfnamefont {E.}~\bibnamefont
  {Florens}}, \bibinfo {author} {\bibfnamefont {O.}~\bibnamefont {Eiff}},\ and\
  \bibinfo {author} {\bibfnamefont {F.~Y.}\ \bibnamefont {Moulin}},\ }\bibfield
   {title} {\bibinfo {title} {Defining the roughness sublayer and its
  turbulence statistics},\ }\href@noop {} {\bibfield  {journal} {\bibinfo
  {journal} {Exp Fluids}\ }\textbf {\bibinfo {volume} {54}},\ \bibinfo {pages}
  {1500} (\bibinfo {year} {2013})}\BibitemShut {NoStop}%
\bibitem [{\citenamefont {Manes}\ \emph {et~al.}(2007)\citenamefont {Manes},
  \citenamefont {Pokrajac},\ and\ \citenamefont {McEwan}}]{manes2007}%
  \BibitemOpen
  \bibfield  {author} {\bibinfo {author} {\bibfnamefont {C.}~\bibnamefont
  {Manes}}, \bibinfo {author} {\bibfnamefont {D.}~\bibnamefont {Pokrajac}},\
  and\ \bibinfo {author} {\bibfnamefont {I.}~\bibnamefont {McEwan}},\
  }\bibfield  {title} {\bibinfo {title} {Double-averaged open-channel flows
  with small relative submergence},\ }\href@noop {} {\bibfield  {journal}
  {\bibinfo  {journal} {J. Hydraul. Eng}\ }\textbf {\bibinfo {volume} {133}},\
  \bibinfo {pages} {896} (\bibinfo {year} {2007})}\BibitemShut {NoStop}%
\bibitem [{\citenamefont {Rouz\`es}\ \emph {et~al.}(2018)\citenamefont
  {Rouz\`es}, \citenamefont {Moulin}, \citenamefont {Florens},\ and\
  \citenamefont {Eiff}}]{rouzes2018}%
  \BibitemOpen
  \bibfield  {author} {\bibinfo {author} {\bibfnamefont {M.}~\bibnamefont
  {Rouz\`es}}, \bibinfo {author} {\bibfnamefont {F.~Y.}\ \bibnamefont
  {Moulin}}, \bibinfo {author} {\bibfnamefont {E.}~\bibnamefont {Florens}},\
  and\ \bibinfo {author} {\bibfnamefont {O.}~\bibnamefont {Eiff}},\ }\bibfield
  {title} {\bibinfo {title} {Low relative-submergence effects in a rough-bed
  open-channel flow},\ }\href@noop {} {\bibfield  {journal} {\bibinfo
  {journal} {J. Hydraul. Res.}\ } (\bibinfo {year} {2018})}\BibitemShut
  {NoStop}%
\bibitem [{\citenamefont {Chagot}\ \emph {et~al.}(2020)\citenamefont {Chagot},
  \citenamefont {Moulin},\ and\ \citenamefont {Eiff}}]{chagot2020}%
  \BibitemOpen
  \bibfield  {author} {\bibinfo {author} {\bibfnamefont {L.}~\bibnamefont
  {Chagot}}, \bibinfo {author} {\bibfnamefont {F.~Y.}\ \bibnamefont {Moulin}},\
  and\ \bibinfo {author} {\bibfnamefont {O.}~\bibnamefont {Eiff}},\ }\bibfield
  {title} {\bibinfo {title} {Towards converged statistics in three-dimensional
  canopy-dominated flows},\ }\href@noop {} {\bibfield  {journal} {\bibinfo
  {journal} {Experiments in Fluids}\ }\textbf {\bibinfo {volume} {24}}
  (\bibinfo {year} {2020})}\BibitemShut {NoStop}%
\bibitem [{\citenamefont {Dupuis}\ \emph {et~al.}(2016)\citenamefont {Dupuis},
  \citenamefont {Proust}, \citenamefont {Berni},\ and\ \citenamefont
  {Paquier}}]{dupuis2016}%
  \BibitemOpen
  \bibfield  {author} {\bibinfo {author} {\bibfnamefont {V.}~\bibnamefont
  {Dupuis}}, \bibinfo {author} {\bibfnamefont {S.}~\bibnamefont {Proust}},
  \bibinfo {author} {\bibfnamefont {C.}~\bibnamefont {Berni}},\ and\ \bibinfo
  {author} {\bibfnamefont {A.}~\bibnamefont {Paquier}},\ }\bibfield  {title}
  {\bibinfo {title} {Combined effects of bed friction and emergent cylinder
  drag in open channel flow},\ }\href@noop {} {\bibfield  {journal} {\bibinfo
  {journal} {Environmental Fluid Mechanics}\ }\textbf {\bibinfo {volume}
  {16}},\ \bibinfo {pages} {1173} (\bibinfo {year} {2016})}\BibitemShut
  {NoStop}%
\bibitem [{\citenamefont {Oukacine}\ \emph {et~al.}(2022)\citenamefont
  {Oukacine}, \citenamefont {Larrarte},\ and\ \citenamefont
  {Goutal}}]{Oukacine2022}%
  \BibitemOpen
  \bibfield  {author} {\bibinfo {author} {\bibfnamefont {M.}~\bibnamefont
  {Oukacine}}, \bibinfo {author} {\bibfnamefont {F.}~\bibnamefont {Larrarte}},\
  and\ \bibinfo {author} {\bibfnamefont {N.}~\bibnamefont {Goutal}},\
  }\bibfield  {title} {\bibinfo {title} {Structure of open-channel flows
  through an array of square cylinders},\ }\href@noop {} {\bibfield  {journal}
  {\bibinfo  {journal} {Urban Water Journal}\ }\textbf {\bibinfo {volume}
  {19}},\ \bibinfo {pages} {732} (\bibinfo {year} {2022})}\BibitemShut
  {NoStop}%
\end{thebibliography}%

\end{document}